\documentclass{article}
\usepackage{spconf,amsmath,epsfig}

\title{Reduced-Rank DOA Estimation based on Joint Iterative Subspace Optimization and Grid Search}
\name{Lei Wang and Rodrigo C. de Lamare}
\address{Communications Research Group, Department of Electronics \\
University of York, York YO10 5DD, UK\\
Email:\{lw517,rcdl500\}@ohm.york.ac.uk}
\begin{document}
\ninept
\maketitle
\begin{abstract}
In this paper, we propose a novel reduced-rank algorithm for
direction of arrival (DOA) estimation based on the minimum variance
(MV) power spectral evaluation. It is suitable to DOA estimation
with large arrays and can be applied to arbitrary array geometries.
The proposed DOA estimation algorithm is formulated as a joint
optimization of a subspace projection matrix and an auxiliary
reduced-rank parameter vector with respect to the MV and grid
search. A constrained least squares method is employed to solve this
joint optimization problem for the output power over the grid. The
proposed algorithm is described for problems of large number of
users' direction finding with or without exact information of the
number of sources, and does not require the singular value
decomposition (SVD). The spatial smoothing (SS) technique is also
employed in the proposed algorithm for dealing with correlated
sources problem. Simulations are conducted with comparisons against
existent algorithms to show the improved performance of the proposed
algorithm in different scenarios.
\end{abstract}
\section{Introduction}

Direction of arrival (DOA) estimation techniques have been widely
employed in many fields related to array processing \cite{Krim}.
Numerous DOA estimation approaches have been considered to date.
Among them are the Capon \cite{Capon}, the conventional
subspace-based methods that require the singular value decomposition
(SVD), such as MUSIC \cite{Schmidt} and ESPRIT \cite{Roy}, and more
recent subspace techniques that do not require the SVD, such as the
auxiliary vector (AV) estimation algorithm \cite{Grover} and the
conjugate gradient (CG) algorithm \cite{Semira}.

The Capon DOA estimation method minimizes the output power of the
undesired interferences while maintaining a constant gain along the
look direction. By computing and plotting Capon's spectrum over the
possible scanning directions, the DOAs can be estimated by locating
the peaks in the spectrum. The estimation accuracy of the Capon
method strongly depends on the number of snapshots and the array
size. The subspace-based MUSIC and ESPRIT algorithms exploit the
eigen-structure of the input covariance matrix to decompose the
observation space into a signal subspace and a corresponding
orthogonal noise subspace. ESPRIT has better performance by
employing a displacement invariance in some specific array
structures. The developed eigen-decomposition algorithms are
described in \cite{Gao}, \cite{Trees}. The previously reported
methods suffer from correlated sources and high computational
complexity due to the eigen-decomposition procedure. The AV and CG
estimation algorithms were proposed recently. The AV method is
developed based on the orthogonality of an extended non-eigenvector
signal subspace with the true signal subspace and the scanning
vector itself. As the scanning vector drops in the signal subspace,
the DOAs are determined by finding the collapse in the extended
signal subspace. The CG method can be considered as an extended
version of the AV method since it applies the residual vectors in
place of the AV basis. Both algorithms show dominate in severe
conditions with a small number of snapshots and at low SNR for both
correlated and uncorrelated sources. However, they work
inefficiently with a large number of sources or without exact
information about the number of sources beforehand.

In this paper, we propose a DOA estimation algorithm by employing a
novel reduced-rank signal processing strategy. The proposed
algorithm is based on a joint iterative subspace optimization (JISO)
and grid search with respect to the MV power spectrum evaluation.
The implementation of the proposed DOA estimation algorithm amounts
to designing a subspace projection matrix and an auxiliary
reduced-rank parameter vector with respect to the MV criterion. We
present a constrained least squares algorithm for jointly estimating
the subspace projection matrix and the auxiliary reduced-rank
parameter vector that calculate the output power over the possible
scanning directions. The proposed algorithm is more practical, in
comparison with the existing algorithms, since it is not limited by
the array structure, does not require the SVD procedure, and works
without information of the source number, which will be shown in
simulations. The estimation accuracy is also satisfied with a large
number of sources' direction finding. In addition, the spatial
smoothing (SS) technique, which was devised by Evans in \cite{Evans}
and further developed by Shan in \cite{Shan}, is employed in the
proposed DOA estimation algorithm for dealing with the problem
caused by correlated sources.

The rest of this paper is organized as follows: we outline a system
model for DOA estimation and present the problem statement in
Section 2. Section 3 derives the proposed DOA estimation algorithm
and analyzes the complexity. The application of the SS technique in
the proposed algorithm is also introduced briefly in this part.
Simulation results are provided and discussed in Section 4, and
conclusions are drawn in Section 5.

\section{System Model and Problem Statement}

\subsection{System Model}

Let us suppose that $q$ narrowband signals impinge on a uniform
linear array (ULA) of $m$ ($m\geq q$) sensor elements. Note that the
proposed DOA estimation algorithm can be applied to arbitrary array
structures. An extension to arbitrary arrays will be sought in a
future work. The ULA here is adopted for using the SS technique and
reaching a fair comparison with ESPRIT, which is applied to some
specific array structures. The $i$th snapshot's vector of sensor
array outputs $\boldsymbol x(i)\in\mathcal C^{m\times 1}$ can be
modeled as
\begin{equation} \label{1}
\centering {\boldsymbol x}(i)={\boldsymbol A}({\boldsymbol
{\theta}}){\boldsymbol s}(i)+{\boldsymbol n}(i),~~~ i=1,\ldots,N
\end{equation}
where
$\boldsymbol{\theta}=[\theta_{0},\ldots,\theta_{q-1}]^{T}\in\mathcal{C}^{q
\times 1}$ is the signal DOAs, ${\boldsymbol A}({\boldsymbol
{\theta}})=[{\boldsymbol a}(\theta_{0}),\ldots,{\boldsymbol
a}(\theta_{q-1})]\in\mathcal{C}^{m \times q}$ is the matrix that
contains the signal direction vectors ${\boldsymbol a}(\theta_{k})$,
where ${\boldsymbol a}(\theta_{k})=[1,e^{-2\pi
j\frac{d}{\lambda_{c}}cos{\theta_{k}}},\ldots,$\\$e^{-2\pi
j(m-1)\frac{d}{\lambda_{c}}cos{\theta_{k}}}]^{T}\in\mathcal{C}^{m
\times 1},~~~(k=0,\ldots,q-1)$, $\lambda_{c}$ is the wavelength, and
$d$ ($d=\lambda_{c}/2$ in general) is the inter-element distance of
the ULA. To avoid mathematical ambiguities, the direction vectors
$\boldsymbol a(\theta_{k})$ are considered to be linearly
independent \cite{Trees}. ${\boldsymbol s}(i)\in
\mathcal{R}^{q\times 1}$ is the source data. ${\boldsymbol
n}(i)\in\mathcal{C}^{m\times 1}$ is the white sensor noise, which is
assumed to be a zero-mean spatially and Gaussian process, $N$ is the
number of snapshots, and $(\cdot)^{T}$\ denotes transpose.

\subsection{Problem statement}
Based on the MV output power spectrum (or Capon output power
spectrum), \cite{Capon}, \cite{Manolakis}, the output power to each
scanning direction for DOA estimation is expressed by
\begin{equation}\label{2}
\begin{split}
&\hat{\theta}=\arg\min_{\theta}{\boldsymbol
w}_{\theta}^{H}{\boldsymbol R}{\boldsymbol w_{\theta}}\\
&\textrm{subject to}~~{\boldsymbol w}_{\theta}^{H}{\boldsymbol
a}(\theta)=1
\end{split}
\end{equation}
where $\hat{\theta}$ is the estimated direction and $\boldsymbol
w_{\theta}=[w_{\theta, 1},\ldots,w_{\theta, m}]^{T}\in\mathcal
C^{m\times 1}$ is the weight vector corresponding to the current
scanning direction $\theta$. $(\cdot)^{H}$ denotes Hermitian
transpose. $\boldsymbol R$ is the data covariance matrix
\begin{equation}\label{3}
\boldsymbol R=E[\boldsymbol x(i)\boldsymbol x^{H}(i)]=\boldsymbol
A(\boldsymbol \theta)\boldsymbol R_{s}\boldsymbol
{A}^{H}(\boldsymbol \theta)+\sigma_{n}^{2}\boldsymbol I
\end{equation}
where $\boldsymbol R_{s}=E[\boldsymbol s(i)\boldsymbol s^{H}(i)]$
denotes the signal covariance matrix, which is diagonal if the
sources are uncorrelated and is nondiagonal and nonsingular for
partially correlated sources, and $E[\boldsymbol n(i)\boldsymbol
n^{H}(i)]=\sigma_{n}^{2}\boldsymbol I$ with $\boldsymbol I$ being
the corresponding identity matrix.

The MV (Capon) power spectrum estimation algorithm attempts to
minimize the contribution of the total output power while
maintaining an unity gain along a look direction $\theta$. By
optimizing the weight vector $\boldsymbol w_{\theta}$ and obtaining
the output power for all possible directions $\theta\in(0^{o},
180^{o})$, the DOAs can be determined by finding the peaks in the
output power spectrum. The weight solution is \cite{Capon},
\cite{Manolakis}
\begin{equation}\label{4}
\boldsymbol w_{\theta}=\frac{\boldsymbol R^{-1}\boldsymbol
a(\theta)}{\boldsymbol a^{H}(\theta)\boldsymbol R^{-1}\boldsymbol
a(\theta)}
\end{equation}

Substituting (\ref{4}) into (\ref{2}), DOA estimation based on the
MV (Capon) power spectrum is given by
\begin{equation}\label{5}
\hat{\theta}_{\textrm{MV}}=\arg\max_{\theta}\big[\boldsymbol
a^{H}(\theta)\boldsymbol R^{-1}\boldsymbol a(\theta)\big]^{-1}
\end{equation}

Note that complete knowledge of $\boldsymbol R$ cannot be obtained
in practice. We may use a sample-average recursion to estimate this
input covariance matrix, which is given by
\begin{equation}\label{6}
\hat{\boldsymbol R} = \frac{1}{N}\sum_{i=1}^{N}\boldsymbol
x(i)\boldsymbol x^H(i)
\end{equation}
Where $\hat{\boldsymbol R}$ is not invertible if the number of
available snapshots is less than the number of sensors ($N\leq m$).
It can be implemented by employing the diagonal loading technique
\cite{Trees}.

The above MV based DOA estimation method suffers from a heavy
computational load for large $m$ due to the matrix inversion and
works inefficiently in the presence of correlated sources.
Furthermore, the performance is inferior when large number of
sources appear in the system.

\section{Proposed DOA Estimation Algorithm}

In this section, we employ a reduced-rank strategy to perform DOA
estimation. This is carried out via the proposed joint iterative
subspace optimization (JISO) according to the MV criterion for
estimating the subspace projection matrix and the auxiliary
reduced-rank parameter vector followed by a grid search.

\subsection{Proposed Reduced-Rank DOA Estimation Scheme}

We introduce a subspace projection matrix $\boldsymbol
T_{r}=[\boldsymbol t_1,~\boldsymbol t_2,~\ldots,~\boldsymbol
t_r]$\\$\in\mathcal C^{m\times r}$, which is responsible for the
dimensionality reduction, to project the $m\times1$ received vector
$\boldsymbol x(i)$ onto a lower dimension, yielding
\begin{equation}\label{7}
\bar{\boldsymbol x}(i)=\boldsymbol T_{r}^{H}\boldsymbol x(i)
\end{equation}
where $\boldsymbol t_l=[t_{1,l}, t_{2,l}, \ldots,
t_{m,l}]^T\in\mathcal C^{m\times1},~(l=1,\ldots,r)$ makes up the
subspace projection matrix $\boldsymbol T_r$, $\bar{\boldsymbol
x}(i)\in\mathcal C^{r\times 1}$ is the projected received vector,
and in what follows, all $r$ dimensional quantities are denoted with
a ``bar". $r<m$ is the rank. An auxiliary filter with the
reduced-rank weight vector $\bar{\boldsymbol
f_{\theta}}=[\bar{f}_{\theta, 1},\bar{f}_{\theta,
2},\ldots,\bar{f}_{\theta, r}]^{T}\in\mathcal C^{r\times 1}$ is
applied after the projection procedure. The aim of $\boldsymbol T_r$
is to extract the key features of the original input vector
$\boldsymbol x(i)$ and form the reduced-rank input vector
$\bar{\boldsymbol x}(i)$. The auxiliary reduced-rank weight vector
$\bar{\boldsymbol f_{\theta}}$ works on $\bar{\boldsymbol x}(i)$ for
obtaining the output power with respect to the current scanning
direction $\theta$. Since the procedure is operated with a lower
dimension $r$, the computational complexity will be reduced if
$r<<m$. Since DOA estimation depends on the number of sensor
elements $m$ and on the eigenvalue spread of the input covariance
matrix, the proposed reduced-rank estimation scheme will exhibit
improved performance under conditions where $m$ is large
\cite{Chen}. Following the MV DOA estimation in (\ref{2}), the
proposed optimization problem can be expressed by
\begin{equation}\label{8}
\begin{split}
&\hat{\theta}_{\textrm{JISO}}=\arg\min_{\theta}\bar{\boldsymbol
f_{\theta}}^{H}{\boldsymbol T}_r^H {\boldsymbol R} {\boldsymbol T}_r \bar{\boldsymbol f_{\theta}}\\
&\textrm{subject to}~~\bar{\boldsymbol f_{\theta}}^{H}{\boldsymbol
T}_r^H {\boldsymbol a}(\theta)=1
\end{split}
\end{equation}
We find that the minimization with respect to (\ref{8}) is
equivalent to the joint optimization of the subspace projection
matrix $\boldsymbol T_r$ and the auxiliary reduced-rank weight
vector $\bar{\boldsymbol f_{\theta}}$. After obtaining $\boldsymbol
T_r$ and $\bar{\boldsymbol f_{\theta}}$, DOA estimation can be
determined by plotting the output power spectrum for the possible
directions and searching for peaks that correspond to the DOAs of
the sources. It is worth noting that, for $r=1$, the novel scheme
becomes a conventional full-rank MV scheme with an additional weight
parameter $\bar{f}_{\theta}$ that provides an amplitude gain. For
$r>1$, the signal processing tasks are changed and $\boldsymbol T_r$
and $\bar{\boldsymbol f_{\theta}}$ are optimized for obtaining the
proposed output power spectrum for the possible directions.

\subsection{Proposed Joint Iterative Subspace Optimization Algorithm}

The challenge left to us is how to efficiently compute the subspace
projection matrix $\boldsymbol T_r$ and the auxiliary reduced-rank
weight vector $\bar{\boldsymbol f_{\theta}}$ for solving the
optimization problem (\ref{8}). We propose a constrained least
squares (LS) algorithm to solve this joint optimization problem. The
constraint in (\ref{8}) can be incorporated by the method of
Lagrange multipliers \cite{Haykin} in the form
\begin{equation}\label{9}
\mathcal {J}=\sum_{l=1}^{i}\alpha^{i-l}\big|\bar{\boldsymbol
f_{\theta}}^{H}(i)\boldsymbol T_{r}^{H}(i)\boldsymbol
x(l)\big|^{2}+\lambda\big[\bar{\boldsymbol
f_{\theta}}^{H}(i)\boldsymbol T_{r}^{H}(i)\boldsymbol
a(\theta)-1\big]
\end{equation}
where $\alpha$ is a forgetting factor, which is a positive constant
close to, but less than $1$, and $\lambda$ is a scalar Lagrange
multiplier. Fixing $\bar{\boldsymbol f_{\theta}}(i)$, computing the
gradient of (\ref{9}) with respect to $\boldsymbol T_{r}(i)$, yields
\begin{equation}\label{10}
\begin{split}
\nabla\mathcal J_{T_r}&=\sum_{l=1}^{i}\alpha^{i-l}\boldsymbol
x(l)\boldsymbol x^{H}(l)\boldsymbol T_{r}(i)\bar{\boldsymbol
f_{\theta}}(i)\bar{\boldsymbol f_{\theta}}^{H}(i)+\lambda_{T_r}\boldsymbol a(\theta)\bar{\boldsymbol f_{\theta}}^{H}(i)\\
&=\hat{{\boldsymbol R}}(i){\boldsymbol T}_{r}(i)\bar{\boldsymbol
f_{\theta}}(i)\bar{\boldsymbol
f_{\theta}}^H(i)+\lambda_{T_r}\boldsymbol a(\theta)\bar{\boldsymbol
f_{\theta}}^{H}(i)
\end{split}
\end{equation}
where $\hat{\boldsymbol R}(i)=\sum_{l=1}^{i}\alpha^{i-l}\boldsymbol
x(l)\boldsymbol x^{H}(l)\in\mathcal C^{m\times m}$ is the estimated
covariance matrix According to \cite{Haykin}, $\hat{\boldsymbol
R}(i)$ can be written in a recursive form as
\begin{equation}\label{11}
\hat{\boldsymbol R}(i)=\alpha\hat{\boldsymbol R}(i-1)+\boldsymbol
x(i)\boldsymbol x^H(i)
\end{equation}

Making the above gradient terms equal to zero, multiplying
$\bar{\boldsymbol f_{\theta}}(i)$ from the right of both sides, and
rearranging the expression, it becomes,
\begin{equation}\label{12}
\boldsymbol T_{r}(i)\bar{\boldsymbol
f_{\theta}}=-\lambda_{T_r}\hat{\boldsymbol R}^{-1}(i)\boldsymbol
a(\theta)
\end{equation}
where $\hat{\boldsymbol R}^{-1}(i)$ is invertible by employing the
diagonal loading technique.

If we define $\hat{\boldsymbol p}(i)=-\lambda_{T_r}\hat{\boldsymbol
R}^{-1}(i)\boldsymbol a(\theta)$, the solution of $\boldsymbol
T_r(i)$ can be regarded to find the solution to the linear equation
\begin{equation}\label{13}
\boldsymbol T_r(i)\bar{\boldsymbol f_{\theta}}=\hat{\boldsymbol
p}(i)
\end{equation}

In order to find an unique solution for $\boldsymbol T_r(i)$, we
express the quantities involved in (\ref{13}) by
\begin{equation}\label{14}
\boldsymbol T_r(i)=\begin{bmatrix}
 \boldsymbol\rho_{1}(i)\\
 \boldsymbol\rho_{2}(i)\\
 \vdots\\
 \boldsymbol\rho_{m}(i)\\ \end{bmatrix};~~
\bar{\boldsymbol f_{\theta}}(i)=\begin{bmatrix}
 \bar{f}_{\theta,1}(i)\\
 \bar{f}_{\theta,2}(i)\\
 \vdots\\
 \bar{f}_{\theta,r}(i)\\ \end{bmatrix};~~
\hat{\boldsymbol p}_{\hat{R}}(i)=\begin{bmatrix}
 \hat{p}_1(i)\\
 \hat{p}_2(i)\\
 \vdots\\
 \hat{p}_m(i)\\ \end{bmatrix}
\end{equation}

The problem in (\ref{13}) is equivalent to find
$\boldsymbol\rho_{j}(i)$ $(j=1, \ldots, m)$ for satisfying
\begin{equation}\label{15}
\min~\|\boldsymbol\rho_j(i)\|^2,~~\textrm{subject~to}~\boldsymbol\rho_j(i)\bar{\boldsymbol
f_{\theta}}=\hat{p}_j(i)
\end{equation}
which is obtained by using the Lagrange multiplier method
\begin{equation}\label{16}
\boldsymbol\rho_j(i)=\hat{p}_j(i)\frac{\bar{\boldsymbol
f_{\theta}}^H(i)}{\|\bar{\boldsymbol f_{\theta}}(i)\|^2}
\end{equation}
and thus the projection matrix is
\begin{equation}\label{17}
\boldsymbol T_r(i)=\hat{\boldsymbol p}(i)\frac{\bar{\boldsymbol
f_{\theta}}^H(i)}{\|\bar{\boldsymbol f_{\theta}}(i)\|^2}
\end{equation}

Substituting the definition of $\hat{\boldsymbol p}(i)$ into
(\ref{17}), we have
\begin{equation}\label{18}
\boldsymbol T_r(i)=-\lambda_{T_r}\hat{\boldsymbol
R}^{-1}(i)\boldsymbol a(\theta)\frac{\bar{\boldsymbol
f_{\theta}}^H(i)}{\|\bar{\boldsymbol f_{\theta}}(i)\|^2}
\end{equation}

The multiplier $\lambda_{T_r}$ can be solved by incorporating
(\ref{12}) with the constraint in (\ref{8}), which is
\begin{equation}\label{19}
\lambda_{T_r}=-\frac{1}{\boldsymbol a^H(\theta)\hat{\boldsymbol
R}^{-1}(i)\boldsymbol a(\theta)}
\end{equation}

Substituting (\ref{19}) into (\ref{18}), we get the projection
matrix
\begin{equation}\label{20}
\boldsymbol T_r(i)=\frac{\hat{\boldsymbol R}^{-1}(i)\boldsymbol
a(\theta)}{\boldsymbol a^H(\theta)\hat{\boldsymbol
R}^{-1}(i)\boldsymbol a(\theta)}\frac{\bar{\boldsymbol
f_{\theta}}^H(i)}{\|\bar{\boldsymbol f_{\theta}}(i)\|^2}
\end{equation}

At the same time, fixing $\boldsymbol T_{r}(i)$, taking the gradient
of (\ref{9}) with respect to $\bar{\boldsymbol f_{\theta}}(i)$, and
making it equal to a null vector, we obtain
\begin{equation}\label{21}
\begin{split}
\nabla\mathcal
J_{\bar{f}_{\theta}}&=\sum_{l=1}^{i}\alpha^{i-l}\boldsymbol
T_{r}^{H}(i)\boldsymbol x(l)\boldsymbol x^{H}(l)\boldsymbol
T_{r}(i)\bar{\boldsymbol f_{\theta}}(i)+\lambda_{\bar{f_{\theta}}}\boldsymbol T_{r}^{H}(i)\boldsymbol a(\theta)\\
&=\hat{\bar{\boldsymbol R}}(i)\bar{\boldsymbol
f_{\theta}}(i)+\lambda_{\bar{f_{\theta}}}\boldsymbol
T_r^H(i){\boldsymbol a}(\theta)
\end{split}
\end{equation}
where $\hat{\bar{\boldsymbol
R}}(i)=\sum_{l=1}^{i}\alpha^{i-l}\bar{\boldsymbol
x}(l)\bar{\boldsymbol x}^{H}(l)\in\mathcal C^{r\times r}$ is the
estimate of the reduced-rank covariance matrix $\bar{\boldsymbol
R}=E[\bar{\boldsymbol x}(i)\bar{\boldsymbol x}^{H}(i)] =\boldsymbol
T_{r}^{H}E[{\boldsymbol x}(i){\boldsymbol x}^{H}(i)]\boldsymbol
T_{r}$.

Following the same procedures for calculating $\boldsymbol T_r(i)$,
we obtain the result for the auxiliary reduced-rank weight vector
$\bar{\boldsymbol f}_{\theta}(i)$
\begin{equation}\label{22}
\bar{\boldsymbol
f_{\theta}}(i)=-\lambda_{\bar{f_{\theta}}}\hat{\bar{\boldsymbol
R}}^{-1}(i)\boldsymbol T_{r}^{H}(i)\boldsymbol a(\theta)
\end{equation}
\begin{equation}\label{23}
\lambda_{\bar{f_{\theta}}}= -\frac{1}{\bar{\boldsymbol
a}^{H}(\theta)\hat{  \bar{\boldsymbol R}}^{-1}(i) \bar{\boldsymbol
a}(\theta) }
\end{equation}
\begin{equation}\label{24}
\bar{\boldsymbol f_{\theta}}(i)=\frac{\hat{\bar{\boldsymbol
R}}^{-1}(i)\bar{\boldsymbol a}(\theta)}{\bar{\boldsymbol
a}^{H}(\theta)\hat{\bar{\boldsymbol R}}^{-1}(i)\bar{\boldsymbol
a}(\theta)}
\end{equation}
where $\bar{\boldsymbol a}(\theta)=\boldsymbol T_{r}^{H}\boldsymbol
a(\theta)\in\mathcal C^{r\times 1}$ is the projected steering vector
with respect to the current scanning direction. Note that (\ref{24})
is similar in form to (\ref{4}) if we do not consider the time
instant $i$. The proposed reduced-rank weight vector
$\bar{\boldsymbol f}_{\theta}(i)$ is more general when dealing with
DOA estimation, namely, for $r=m$, it is equivalent to the MV weight
vector, and, for $1<r<m$, it operates under lower dimensions for
reducing the complexity and improving the performance.
\begin{table}[t]
\centering
    \caption{The JISO algorithm for each
scanning direction}
    \label{tab:The JISO algorithm for each
scanning direction}
    \begin{small}
    \begin{tabular}{l}
\hline
\bfseries {Initialization:}\\
$\boldsymbol T_{r}(0)=[\boldsymbol I_{r}^{T}~ \boldsymbol 0_{r\times
(m-r)}^{T}]$\\
$\bar{\boldsymbol f_{\theta}}(0)=\big(\boldsymbol
T_{r}^{H}(0)\boldsymbol a(\theta_{n})\big)/\big(\|\boldsymbol
T_{r}^{H}(0)\boldsymbol a(\theta_{n})\|^{2}\big)$\\
\bfseries {Update for each time instant} $i=1,\ldots,N$\\
$\bar{\boldsymbol x}(i)=\boldsymbol T_{r}^{H}(i-1)\boldsymbol x(i)$\\
$\bar{\boldsymbol a}(\theta_{n})=\boldsymbol
T_{r}^{H}(i-1)\boldsymbol a(\theta_{n})$\\
$\hat{{\boldsymbol R}}(i)=\alpha\hat{{\boldsymbol
R}}(i-1)+{\boldsymbol x}(i){\boldsymbol x}^{H}(i)$\\
$\hat{\bar{\boldsymbol R}}(i)=\alpha\hat{\bar{\boldsymbol
R}}(i-1)+\bar{\boldsymbol x}(i)\bar{\boldsymbol x}^{H}(i)$\\
$\bar{\boldsymbol f_{\theta}}(i)={\hat{\bar{\boldsymbol
R}}^{-1}(i)\bar{\boldsymbol a}(\theta)}/\big({\bar{\boldsymbol
a}^{H}(\theta)\hat{\bar{\boldsymbol R}}^{-1}(i)\bar{\boldsymbol
a}(\theta)}\big)$\\
$\boldsymbol T_r(i)=\frac{\hat{\boldsymbol R}^{-1}(i)\boldsymbol
a(\theta)}{\boldsymbol a^H(\theta)\hat{\boldsymbol
R}^{-1}(i)\boldsymbol a(\theta)}\frac{\bar{\boldsymbol
f_{\theta}}^H(i)}{\|\bar{\boldsymbol f_{\theta}}(i)\|^2}$\\
\bfseries {Output power}\\
$P_{\textrm{JISO}}(\theta_{n})=1/\big(\bar{\boldsymbol
a}^{H}(\theta_{n})\hat{\bar{\boldsymbol R}}^{-1}\bar{\boldsymbol
a}(\theta_{n})\big)$\\
\hline
    \end{tabular}
    \end{small}
\end{table}

\subsection{DOA Estimation}

After $N$ snapshots, substituting the weight solution
$\bar{\boldsymbol f}_{\theta}$ expressed in (\ref{24}) with respect
to the possible scanning directions $\theta\in(0^{o}, 180^{o})$, and
the subspace projection matrix $\boldsymbol T_r$ in (\ref{20}) into
(\ref{8}), we obtain the corresponding output power spectrum for DOA
estimation
\begin{equation}\label{25}
P_{\textrm{JISO}}(\theta_{n})=\big(\bar{\boldsymbol
a}^{H}(\theta_{n})\hat{\bar{\boldsymbol R}}^{-1}\bar{\boldsymbol
a}(\theta_{n})\big)^{-1}
\end{equation}
where the scanning direction $\theta_{n}=n\Delta^{o}$, $\Delta^{o}$
is the search step, and $n=1,2,\ldots,180^{o}/\Delta^{o}$. For a
simple and convenient search, we make $180^{o}/\triangle^{o}$ an
integer. We use a similar form to that of (\ref{11}) for estimating
$\hat{\bar{\boldsymbol R}}$. The proposed JISO algorithm for each
scanning direction $\theta_{n}$ is summarized in Table \ref{tab:The
JISO algorithm for each scanning direction}, where $\boldsymbol
T_{r}(0)$ and $\bar{\boldsymbol f_{\theta}}(0)$ are initialized to
ensure the constraint. The proposed algorithm provides an iterative
exchange of information between the projection matrix and the
reduced-rank weight vector, which leads to the improved performance.

The output power in (\ref{25}) is much higher if the scanning
direction $\theta_{n}=\theta_{k}, ~(k=0,\ldots,q-1)$, which
corresponds to the transmitted sources, compared with other scanning
angles that correspond to the noise level. Therefore, the output
power spectrum shows peaks with respect to the sources when we plot
it through the whole search range.


Considering correlated sources, we can use the SS technique
\cite{Evans} in our proposed algorithm. It is based on averaging the
covariance matrix of identical overlapping arrays and so requires an
array of identical elements equipped with some form of periodic
structure. We divide the ULA into overlapping subarrays of size $n$,
with elements $\{1, \ldots, n\}$ forming the first subarray,
elements $\{2, \ldots, n+1\}$ forming the second subarray, etc., and
$J=m-n+1$ as the number of subarrays. Note that the selection of $n$
needs to follow $n\geq q$ and $J\geq q$ \cite{Shan}. The SS
preprocessing scheme operates on the input vector $\boldsymbol x(i)$
to obtain each subarray vector $\boldsymbol x_j(i)$, where
$j=1,\ldots,J$. The proposed JISO algorithm is followed for DOA
estimation. We denominate this SS-based algorithm as JISO-SS, which
is summarized in Table \ref{tab:The JISO-SS algorithm for each
scanning direction}, where $\boldsymbol T_{r,\textrm{ss}}\in\mathcal
C^{n\times r}$ and $\bar{\boldsymbol
f}_{\theta,\textrm{ss}}\in\mathcal C^{r\times 1}$ are the SS-based
subspace projection matrix and auxiliary reduced-rank weight vector,
respectively. $\boldsymbol A_{\textrm{ss}}=[\boldsymbol
a_{\textrm{ss}}(\theta_0), \ldots, \boldsymbol
a_{\textrm{ss}}(\theta_{q-1})]\in\mathcal C^{n\times q}$ is the
matrix that contains the direction vectors $\boldsymbol
a_{\textrm{ss}}(\theta_{k})$, where $\boldsymbol
a_{\textrm{ss}}(\theta_{k})=[1, e^{-2\pi
j\frac{d}{\lambda_c}}\cos\theta_{k}, \ldots, e^{-2\pi
j(n-1)\frac{d}{\lambda_c}}\cos\theta_{k}]^T\in\mathcal C^{n\times
1}, (k=0, \ldots, q-1)$, $\boldsymbol D=\textrm{diag}\{e^{-2\pi
j\frac{d}{\lambda_c}\cos\theta_0}, \ldots, e^{-2\pi
j\frac{d}{\lambda_c}\cos\theta_{q-1}}\}$\\$\in\mathcal C^{q\times
q}$, and the subscript ``ss" denotes that it is for the SS-based
proposed algorithm. $\boldsymbol x_{j,\textrm{ss}}(i)\in\mathcal
C^{n\times 1}$ is the input vector at the $j$th subarray,
$\bar{\boldsymbol x}_{j,\textrm{ss}}(i)\in\mathcal C^{r\times 1}$ is
the corresponding reduced-rank input vector, and $\boldsymbol
n_{j,\textrm{ss}}(i)\in\mathcal C^{n\times 1}$ is the white sensor
noise. $\hat{\boldsymbol P}_{j,\textrm{ss}}(i)\in\mathcal C^{n\times
n}$ and $\hat{\bar{\boldsymbol P}}_{j,\textrm{ss}}(i)\in\mathcal
C^{r\times r}$ are the full-rank and reduced-rank covariance
matrices of the $j$th subarray, respectively, at time instant $i$.
Since it is a well-known technique for dealing with correlated
sources, the details have been omitted but related references can be
found in \cite{Evans}, \cite{Shan}.
\begin{table}[t]
\centering
    \caption{The JISO-SS algorithm for each
scanning direction}
    \label{tab:The JISO-SS algorithm for each
scanning direction}
    \begin{small}
    \begin{tabular}{l}
\hline
\bfseries {Initialization:}\\
$\boldsymbol T_{r,\textrm{ss}}(0)=[\boldsymbol I_{r}^{T}~
\boldsymbol 0_{r\times
(n-r)}^{T}]$\\
$\bar{\boldsymbol f}_{\theta,\textrm{ss}}(0)=\big(\boldsymbol
T_{r,\textrm{ss}}^{H}(0)\boldsymbol
a_{\textrm{ss}}(\theta_{n})\big)/\big(\|\boldsymbol
T_{r,\textrm{ss}}^{H}(0)\boldsymbol a_{\textrm{ss}}(\theta_{n})\|^{2}\big)$\\
\bfseries {Update for each time instant} $i=1,\ldots,N$\\
~~~~~~{\bfseries for} $j=1, \ldots, J$\\
~~~~~~$\boldsymbol x_{j,\textrm{ss}}(i)=\boldsymbol
A_{\textrm{ss}}\boldsymbol
D^{j-1}\boldsymbol s(i)+\boldsymbol n_{j,\textrm{ss}}(i)$\\
~~~~~~$\bar{\boldsymbol x}_{j,\textrm{ss}}(i)=\boldsymbol T_{r,\textrm{ss}}^{H}(i-1)\boldsymbol x_{j,\textrm{ss}}(i)$\\
~~~~~~$\hat{\boldsymbol P}_{j,\textrm{ss}}(i)=\boldsymbol x_{j,\textrm{ss}}(i)\boldsymbol x_{j,\textrm{ss}}^H(i)$\\
~~~~~~$\hat{\bar{\boldsymbol P}}_{j,\textrm{ss}}(i)=\bar{\boldsymbol x}_{j,\textrm{ss}}(i)\bar{\boldsymbol x}_{j,\textrm{ss}}^H(i)$\\
~~~~~~{\bfseries end}\\
$\bar{\boldsymbol a}_{\textrm{ss}}(\theta_{n})=\boldsymbol
T_{r,\textrm{ss}}^{H}(i-1)\boldsymbol a_{\textrm{ss}}(\theta_{n})$\\
$\hat{{\boldsymbol R}}_{\textrm{ss}}(i)=\alpha\hat{{\boldsymbol
R}}_{\textrm{ss}}(i-1)+\frac{1}{J}\sum_{j=1}^{J}\hat{\boldsymbol P}_{j,\textrm{ss}}(i)$\\
$\hat{\bar{\boldsymbol
R}}_{\textrm{ss}}(i)=\alpha\hat{\bar{\boldsymbol
R}}_{\textrm{ss}}(i-1)+\frac{1}{J}\sum_{j=1}^{J}\hat{\bar{\boldsymbol P}}_{j,\textrm{ss}}(i)$\\
$\bar{\boldsymbol f}_{\theta,\textrm{ss}}(i)={\hat{\bar{\boldsymbol
R}}_{\textrm{ss}}^{-1}(i)\bar{\boldsymbol
a}_{\textrm{ss}}(\theta_{n})}/\big({\bar{\boldsymbol
a}_{\textrm{ss}}^{H}(\theta_{n})\hat{\bar{\boldsymbol
R}}_{\textrm{ss}}^{-1}(i)\bar{\boldsymbol
a}_{\textrm{ss}}(\theta_{n})}\big)$\\
$\boldsymbol T_{r,\textrm{ss}}(i)=\frac{\hat{\boldsymbol
R}_{\textrm{ss}}^{-1}(i)\boldsymbol
a_{\textrm{ss}}(\theta_n)}{\boldsymbol
a_{\textrm{ss}}^H(\theta_n)\hat{\boldsymbol
R}_{\textrm{ss}}^{-1}(i)\boldsymbol
a_{\textrm{ss}}(\theta_n)}\frac{\bar{\boldsymbol
f}_{\theta_n}^H(i)}{\|\bar{\boldsymbol f}_{\theta_n}(i)\|^2}$\\
\bfseries {Output power}\\
$P_{\textrm{JISO},\textrm{ss}}(\theta_{n})=1/\big(\bar{\boldsymbol
a}_{\textrm{ss}}^{H}(\theta_{n})\hat{\bar{\boldsymbol
R}}_{\textrm{ss}}^{-1}\bar{\boldsymbol
a}_{\textrm{ss}}(\theta_{n})\big)$\\
\hline
    \end{tabular}
    \end{small}
\end{table}

Checking the computational complexity, the conventional Capon
\cite{Capon}, MUSIC \cite{Schmidt} and ESPRIT \cite{Roy} algorithms
work with $O(m^{3})$, and the recent AV \cite{Grover} and CG
algorithms have a higher computational cost \cite{Semira}. With
respect to the proposed algorithm, $\hat{\boldsymbol R}^{-1}(i)$
costs $O(m^{3})$ but is invariable for the grid search, namely, the
result obtained for the first scanning direction can be used for the
rest. The complexity of the proposed JISO algorithm for each
iteration is $O(r^{3})$, which is less complex than the AV or CG
methods if $r<<m$ for large arrays. The complexity of the proposed
JISO algorithm with constrained LS optimization method is slightly
higher than the MUSIC and ESPRIT methods and lower than the AV and
CG algorithms. The JISO-SS algorithm is marginally more complex than
the JISO one due to the SS preprocessing. Actually, we can employ
other methods to solve the joint optimization problem (e.g.,
stochastic gradient, recursive least squares \cite{Haykin}) to avoid
the matrix inversion for complexity reduction, which will be
analyzed in the near future.

\section{Simulations}
Simulations are performed for an ULA with half wavelength
interelement spacing. We compare the proposed algorithm with the
Capon, MUSIC, ESPRIT, AV, and CG methods, and run $K=1000$
iterations to get each curve. The SS technique is employed for each
algorithm to improve the performance. In all experiments, the BPSK
signals' power is $\sigma_{s}^{2}=1$ and the noise is spatially and
temporally white Gaussian. The search step is $\Delta^{o}=1^{o}$.
The DOAs are considered to be resolved if
$|\hat{\theta}_{\textrm{JISO}}-\theta_{k}|<1^{o}$.

\begin{figure}[!htb]
\begin{center}
\def\epsfsize#1#2{1.1\columnwidth}
\epsfbox{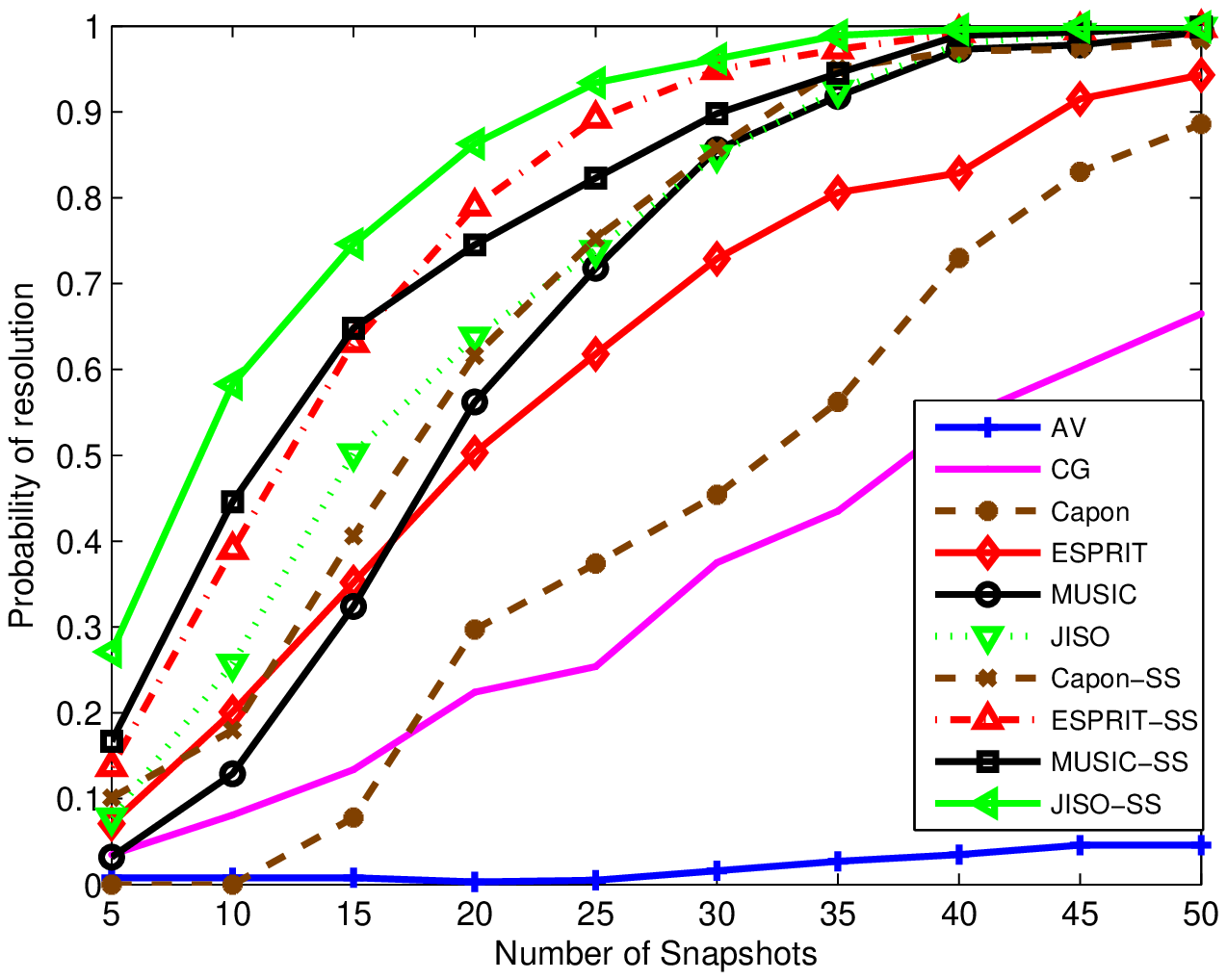} \caption{Probability of resolution versus number
of snapshots (separation $3^{o}$, SNR$=-2$dB, $q=2$, c$=0.9$,
$m=30$, $r=6$, $\delta=5\times10^{-4}$, $\alpha=0.998$, $n=26$)}
\label{fig:snap_snr_2users_highcor_final1}
\end{center}
\end{figure}

In Fig. \ref{fig:snap_snr_2users_highcor_final1}, we consider the
presence of $q=2$ highly correlated sources separated by $3^{o}$
with correlation value $c=0.9$, which are generated as follows:
\begin{equation}\label{26}
s_{1}\sim \mathcal
{N}(0,\sigma_{s}^{2})~~~and~~~s_{2}=cs_{1}+\sqrt{1-c^{2}}s_{3}
\end{equation}
where $s_{3}\sim\mathcal {N}(0,\sigma_{s}^{2})$. The sensor elements
number is $m=30$ and input SNR $=-2$dB. We set the forgetting factor
$\alpha=0.998$, the reduced dimension $r=6$, and the diagonal
loading $\delta=5\times10^{-4}$ for the covariance matrix inverse in
(\ref{18}) and (\ref{19}). The probability of resolution
\cite{Grover}, \cite{Semira} is plotted against the number of
snapshots. The proposed algorithm outperforms other existing methods
with small number of snapshots. The curves between the proposed and
the MUSIC algorithms are shown to intersect when the number of
snapshots increases. The performance of the AV and CG methods can be
seen to be inferior when compared to the proposed algorithm for all
observation periods. Regarding the SS-based algorithms, we set the
subarray size to $n=26$, which accords with \cite{Shan} and reaches
a high probability of resolution. The performance of the algorithms
with the SS technique is improved and the proposed algorithm still
has better performance than the existing ones.

\begin{figure}[!htb]
\begin{center}
\def\epsfsize#1#2{1.1\columnwidth}
\epsfbox{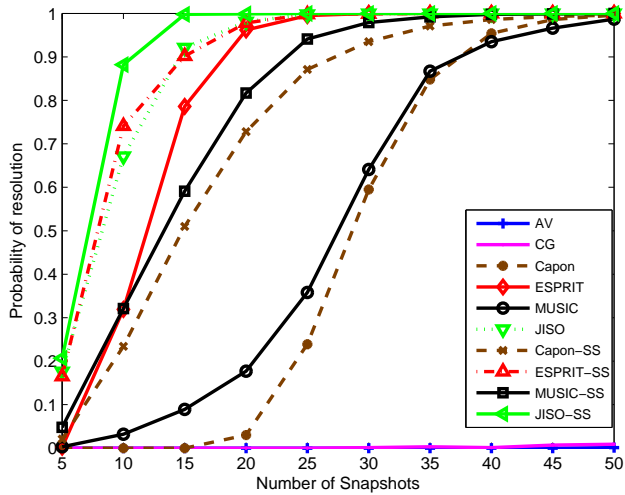} \caption{Probability of resolution versus number
of snapshots (separation $3^{o}$, SNR$=-5$dB, q$=10$, $m=50$, $r=6$,
$\delta=5\times10^{-4}$, $\alpha=0.998$, $n=41$).}
\label{fig:snap_snr_moreusers_final1}
\end{center}
\end{figure}

Next, we consider the sources to be uncorrelated but increase the
number of sources by setting $q=10$. The input SNR $=-5$dB and the
number of sensor elements is set to $m=50$. As can be seen in Fig.
\ref{fig:snap_snr_moreusers_final1}, the AV and CG methods are
unable to obtain a DOA estimate with a large number of sources. The
proposed algorithm demonstrates an improved performance and is the
first to reach the highest resolution, as compared with the
conventional Capon and the subspace-based MUSIC and ESPRIT methods,
following the increase of number of snapshots. The subarray size is
$n=41$ in this scenario.

\begin{figure}[!htb]
\begin{center}
\def\epsfsize#1#2{1.1\columnwidth}
\epsfbox{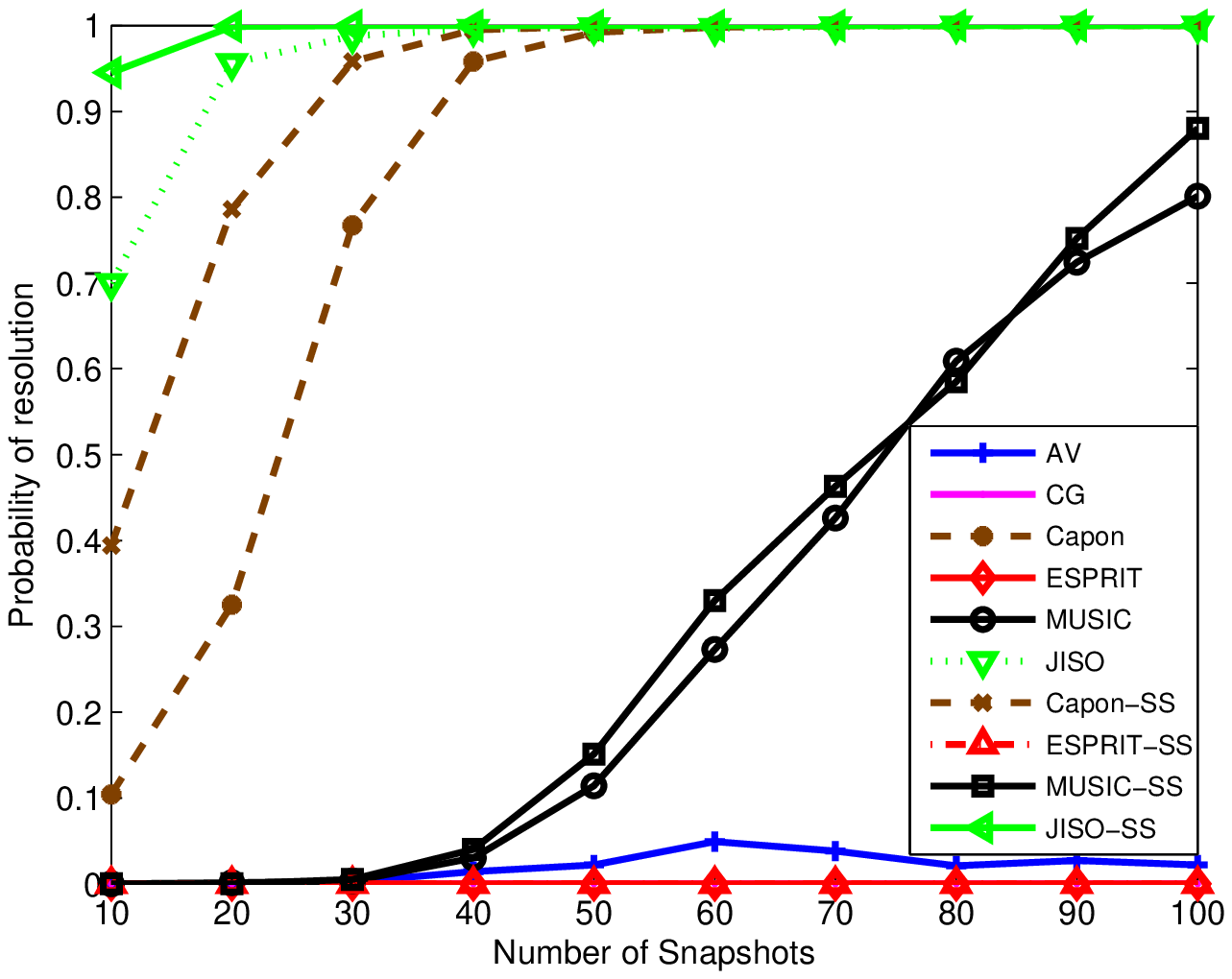} \caption{Probability of resolution versus number
of snapshots (separation $3^{o}$, SNR$=0$dB, $q_{w}=9$, $m=50$,
$r=6$, $\delta=5\times10^{-4}$, $\alpha=0.998$, $n=41$).}
\label{fig:snap_snr_moreusers_wrongq_final1}
\end{center}
\end{figure}

In the last experiment, we assess the performance of the proposed
and analyzed algorithms with an uncorrect number of sources
$q_{w}\neq q$ known by the receiver. This is more practical since
the exact sources number has to be determined by procedures with
extra computation cost and time. We keep the scenario as that in
Fig. \ref{fig:snap_snr_moreusers_final1} but assume an uncorrect
number of sources $q_{w}=9$ instead of $q=10$, and increase the
number of snapshots for the operation. The fixed input SNR $=0$dB.
In Fig. \ref{fig:snap_snr_moreusers_wrongq_final1}, the MUSIC and
its SS-based algorithms start to work with large number of
snapshots, and the ESPRIT and its SS-based algorithms fail to
resolve DOA estimation with the increase of the snapshots since $q$
is critical to the eigendecomposition for the partition of the
signal subspace and the noise subspace in the input covariance
matrix. Also, the design of the AV basis and CG residual vectors
depends strongly on $q$. The Capon and its SS-based algorithms work
well under this condition since they are insensitive to the number
of sources. The same holds for the proposed and its SS-based
algorithms, but both exhibit better performance and lower
complexity. We consider $q_{w}>q$ condition and get the same result.

\section{Concluding Remarks}
We proposed a novel reduced-rank strategy to implement joint
iterative subspace optimization and grid search for DOA estimation.
The DOA estimation problem is formulated as a reduced-rank MV
optimization problem. A subspace projection matrix is introduced to
obtain the covariance matrix processed in the lower dimension so
that computation cost is reduced and performance improved. An
auxiliary reduced-rank parameter vector is combined to realize the
joint iterative optimization with respect to the MV output power for
each scanning direction. By searching the possible directions, the
DOAs can be determined by finding the peaks in the output power
spectrum. The proposed DOA estimation algorithm demonstrates
advantages under large array condition with uncorrelated or
correlated sources. Its performance is not significantly influenced
by some parameters (e.g., the number of sources). In future work, we
will provide the analysis including the Cramer-Rao Bound (CRB) and
compare it with the estimation accuracy of the proposed algorithm.
We will also consider unitary versions of the proposed reduced-rank
algorithm for ULA that do not reuqire grid search.

\end{document}